\documentclass[a4paper,onecolumn]{IEEEtran}
\usepackage{amsmath,amsfonts,amssymb,graphicx}

\begin{document}

\author{Bernd G\"unther\thanks{Beilstein-Institut,
Trakehner Stra\ss e 7-9, 60487 Frankfurt, Germany.}}

\IEEEspecialpapernotice{This paper has been published with copyright by IEEE\\
IEEE Trans. Inf. Theory, 54(7):3206--3210, 2008; DOI 10.1109/TIT.2008.924658.}

\title{On the Probability Distribution of Superimposed Random Codes}

\maketitle


\begin{abstract}
A systematic study of the probability distribution of superimposed random
codes is presented through the use of generating functions. Special attention
is paid to the cases of either uniformly distributed but not necessarily
independent or non uniform but independent bit structures. Recommendations for
optimal coding strategies are derived.
\end{abstract}


\newtheorem{theorem}{Theorem}
\newtheorem{definition}{Definition}

\section{Introduction}
Chemical structure retrieval systems are frequently presented with the task to
produce a list of all stored chemical graphs containing a prescribed
subgraph\cite{Bar93,BarWal98}.
Due to the absence of a linear order among the stored data tree
based search strategies fail, and a sequential search has to be performed. To
accelerate this time consuming process, the actual graph theoretical
substructure match is preceeded by \emph{prescreening}: the entire database is
matched against a library of simple but common \emph{descriptors}, and the
validity of descriptors is recorded in a bitstring for each stored
structure. Suitable choices for descriptors are small chemical subgraphs
containing only few vertices, graph diameters, ring sizes or any other
property that passes from subgraphs to supergraphs. When a query structure is
submitted to the system, the descriptors are evaluated for this query
structure resulting in a query bit string.
Only those stored structures are candidates for a match where
each bit is turned on in all those positions where the query bits are turned
on and will be subjected to the expensive graph theoretical matching
algorithm.

For example, let us consider the compounds in table~\ref{tbl1}.
A chemist might ask for a list of all structures in our database containing
2-(cyclohexylmethyl)naphthalene, which is too complex to be one of the index
descriptors. However, any matching structure must necessarily contain
cyclohexane and naphthalene, and these might be indexed.
We will produce an intermediate result set that also contains
2-(2-cyclohexylethyl)naphthalene, which is not in accordance with the original
query specification and must be singled out by graph matching.
\begin{table}\label{tbl1}
\begin{center}
\begin{tabular}{cc}
\multicolumn{2}{c}{\begin{minipage}{50mm}\begin{center}
\includegraphics{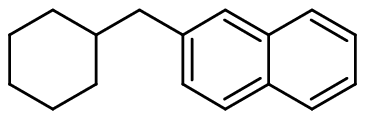}\\
2-(cyclohexylmethyl)naphthalene
\end{center}\end{minipage}}\vspace{2mm}\\
\begin{minipage}{20mm}\begin{center}
\includegraphics{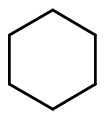}\\
cyclohexane
\end{center}\end{minipage}
&\begin{minipage}{30mm}\begin{center}
\includegraphics{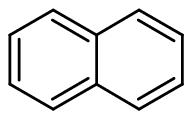}\\
naphthalene
\end{center}\end{minipage}\\
\multicolumn{2}{c}{\begin{minipage}{50mm}\begin{center}
\includegraphics{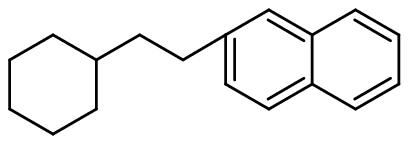}\\
2-(2-cyclohexylethyl)naphthalene
\end{center}\end{minipage}}
\end{tabular}
\end{center}
\caption{}
\end{table}

The Beilstein database of organic compounds contains around 10 million
structures, each a chemical graph of up to 255 vertices,
and the number of descriptors will have the magnitude of one
thousand. With such characteristics, the preevaluated bitstrings will consume
a considerable amount of storage and hence of processing time. On the other
hand, one may expect the $1$-bits to be relatively scarce, whence it should be
possible to compress the bitstrings without losing too much information.
Thus, we are looking for a map $\psi:\dot{I}^N\to\dot{I}^n$,
$\dot{I}=\{0,1\}$ transforming bitstrings of length $N$ into strings of length
$n$. However, we have to observe the partial order relation
$\beta\le\beta'$ between bitstrings defined such that the bits should satisfy
$\beta(i)\le\beta'(i)$ at all positions $i$.
Since we want to use the compressed strings in the same manner 
(by bitmasking) as the
original ones the transformation must be monotone: $\beta\le\beta'\Rightarrow
\psi(\beta)\le\psi\left(\beta'\right)$ and in particular
$\psi(\beta)\vee\psi\left(\beta'\right)\le
\psi\left(\beta\vee\beta'\right)$, where the wedge denotes the bitwise
inclusive or operation.
If the latter inequality was strict
information would be lost, hence we claim:
\begin{equation}
\psi(\beta)\vee\psi\left(\beta'\right)=
\psi\left(\beta\vee\beta'\right)
\end{equation}
for any two source strings $\beta,\beta'\in\dot{I}^N$. This is the defining
condition of superimposed coding.
If we denote by $\beta_j\in\dot{I}^N$ the elementary string having bit $1$ in
position $j$ and $0$ elsewhere and set
$\psi_j:=\psi\left(\beta_j\right)\in\dot{I}^n$ equal to the code word assigned
to bit $j$, then we must have
\begin{equation}\label{eqsupcod}
\psi(\beta)=\bigvee_{j\in\beta^{-1}(1)}\psi_j,
\end{equation}
explaining the term.
Here $\beta^{-1}(1)$ is the inverse image of $1$, i.e.\ the set of all
positions $i$ where the $i$-th bit $\beta(i)$ is turned \emph{on}.
As summary we emphasize: in our context superposition is not a matter of
choice but a requirement.

We speak of \emph{superimposed random coding} if the code words $\psi_j$ are
constructed using a random number generator. Non random approaches have been
considered e.g.\ in \cite{KauS64,Rash90} and perform very
well when the number of
code words superposed in equation~(\ref{eqsupcod}) is bounded, but are
unpredictable beyond this barrier. The random approach was probably initiated
by \cite{Moo48}, but there an invalid probability analysis was given. A
thorough study of the case where source bitstrings and code words are of fixed
weight was given in \cite{Rob79}.
Bloom filters (cf.\ \cite{Blo70} and \cite[p.572]{Knu3}) constitute an
application of this technique where the primary keys of a database are encoded
in a bitstring; it should be noted that the approach favored by us produces a
twodimensional bit array for the entire database, one bitstring for each
record.
Despite the continuing popularity of the
subject in context of chemical substructure search a broader systematic
approach seems to be lacking, a gap that is intended to be filled by the
current paper. 
One question that has to be addressed is the optimal choice of codes.
It is known at least since Roberts' paper \cite{Rob79} that
the target bitstrings should contain a more or less even balance of $0$ and
$1$-bits, but if we want to achieve this other than by try and error we must
study the distribution of target bits.
Of particular practical importance is the situation where the
source bits are turned on with non uniform probability. A few statements
hereabout are by Roberts in \cite{Rob79} but are based on intuition,
not computation, and in fact we disagree with his conclusion. In \cite{Hod76}
the author gives recommendations about the selection of descriptors,
whereas we set ourselves the task
of adapting the coding design to a given set of descriptors.

We must notice that the notion ``random coding'' has an inherent semantic
difficulty. It must be understood clearly that coding and decoding of a bit
pattern have to be performed within one and the same run of the code
generation random experiment. After each coding-decoding cycle the code
generation should be
repeated independently. However, this procedure does not exactly
fit when recording bit patterns in databases: here the codes must be fixed
once and for all. Since a large set of $N$ codes is needed, we might expect
good statistical behavior in this context too.

The observant reader will have noticed that equation~(\ref{eqsupcod})
allows for a target string $\psi(\beta)$ to be covered by a comparison
pattern, $\psi(\beta)\le\psi(\gamma)$, even if the source patterns do not
cover: $\beta\not\le\gamma$.
This will result in the inclusion of fake hits in our result sets.
As long as their number is small this is acceptable, because the bit
comparison will be followed by graph matching anyway.
However, we want to predict and minimize this number.

In section~\ref{secIsotr} we will develop the probabilistic tools necessary to
manage our random bit strings. 
In section~\ref{secfwc} we demonstrate how to compute the distribution of the
target bits from those of source and code bits, and
binomially distributed and fixed weight code
words are introduced as primary examples for code generation.
This settles the case of uniform but not necessarily independent source
bit distributions.
The requirement of uniformity will be dropped in section~\ref{secNuInd} but
the assumption of independence will be added.
The completely general case will be addressed in section~\ref{secgencas}.

\section{Isotropic bit distributions}\label{secIsotr}
Let's consider a fixed source bit pattern $\beta\in\dot{I}^N$ and a fixed
target bit pattern $\alpha\in\dot{I}^n$; then
\begin{eqnarray}
P\left(\psi(\beta)\le\alpha\right)
&=&P\left(\forall j\in\beta^{-1}(1):\psi_j\le\alpha\right)\\
&=&\prod_{j\in\beta^{-1}(1)}P\left(\psi_j\le\alpha\right),
\end{eqnarray}
where the probability is that of the code generation. Now varying the source
pattern $\beta$ independently we obtain an expression for the probability of
target patterns:
\begin{equation}\label{eqtardis}
P\left(\psi(-)\le\alpha\right)=
\sum_{\beta\in\dot{I}^N}P(\beta)
\prod_{j\in\beta^{-1}(1)}P\left(\psi_j\le\alpha\right).
\end{equation}

\begin{definition}
We call a probability distribution on $\dot{I}^n$ \emph{iso\-tropic}, if it
depends only on the number of $1$-bits but not on their position.
\end{definition}

This means that the distribution is invariant under all coordinate
permutations (In \cite{Ren65} the term homogeneous is used).
As is evident from equation~(\ref{eqtardis}), the distribution
of the target patterns is isotropic if the code generation is, even if the
source patterns are non isotropically distributed. Since this observation
leads to a considerable simplification of our analysis we now give a short
exposition of isotropic distributions.

We are given numbers $p_0,\ldots p_n\ge0$ with
$\sum_{k=0}^n\binom{n}{k}p_k=1$
such that the probability of a particular bit pattern $\alpha\in\dot{I}^n$ is
given by $P(\alpha)=p_a$ with $a:=\#\alpha^{-1}(1)$. Our main analytical tool
will be the probability generating function
\begin{equation}
f(t):=\sum_{k=0}^n\binom{n}{k}p_kt^k,
\end{equation}
this being the standard definition obtained by considering the number of
$1$-bits as random variable. For fixed $\alpha\in\dot{I}^n$ with
$a:=\#\alpha^{-1}(1)$ we define
\begin{eqnarray}
F_a&:=&P\left(\xi\le\alpha\right)=\sum_{k=0}^a\binom{a}{k}p_k\label{eqcoef}\\
G_a&:=&P\left(\xi\ge\alpha\right)=\sum_{k=a}^n\binom{n-a}{k-a}p_k.
\end{eqnarray}
Of course the quantities $F_a$ and $G_{n-a}$ are dual to each other, in fact
the one is transformed into the other by switching $0$ and $1$-bits. They play
very different roles for us, though: notice the occurrence of $F_a$ on the
right hand side of equation~(\ref{eqtardis}). $G_a$ is the relative number of
candidates that will be selected by prescreening with bit pattern $\alpha$,
because the condition $\xi\ge\alpha$ singles out precisely those patterns
$\xi$ that have $1$-bits in the same positions as $\alpha$, and probably some
more.
We define generating functions as follows:
\begin{eqnarray}
F(t)&:=&\sum_{m=0}^n\binom{n}{m}F_mt^m\\
G(t)&:=&\sum_{k=0}^n\binom{n}{k}G_{n-k}t^k
\end{eqnarray}
The following relations are readily established:
\begin{eqnarray}
F(t)&=&(1+t)^nf\left(\frac{t}{1+t}\right)\label{eqinvdis}\\
G(t)&=&(-1)^nF(-1-t)\\
G(t)&=&t^nf\left(\frac{1+t}{t}\right)\label{eqf2G}\\
p_m&=&\sum_{k=0}^m(-1)^{m+k}\binom{m}{k}F_k\\
G_m&=&\sum_{k=0}^m(-1)^k\binom{m}{k}F_{n-k}
\end{eqnarray}
The three sets of coefficients $p_m$, $F_m$ and $G_m$ carry the same
information and can be readily converted into each other using the above
relations. The reader will certainly notice the difference between
(\ref{eqinvdis}) and the standard situation \cite[Thm.~XI.1]{Fel70}, which
arises from the fact that the values
$\alpha$ appearing in definition (\ref{eqcoef}) are partially but not linearly
ordered.

For later reference we need the derivatives of $f$ at $1$. By (\ref{eqf2G}) we
have
$f(1+\varepsilon)=\varepsilon^nG\left(\frac{1}{\varepsilon}\right)
=\sum_{k=0}^n\binom{n}{k}G_k\varepsilon^k$.
This allows to read off the Taylor coefficients of the function $f$
at the point $1$ at once:
\begin{equation}\label{eqfderiv}
\frac{f^{(k)}(1)}{k!}=\binom{n}{k}G_k
\end{equation}

We define the moments $\mu_m=\sum_{k=0}^nk^m\binom{n}{k}p_k$
of our distribution as usual and remember their generating
function\cite[Exc.~XI.7.24]{Fel70}:
\begin{equation}\label{eqmomgen}
f\left(e^t\right) =
\sum_{m=0}^\infty \frac{\mu_m}{m!}t^m.
\end{equation}
Equations~(\ref{eqfderiv}) and (\ref{eqmomgen}) allow to express the first two
moments in terms of the distribution coefficients $F_m$:
\begin{align}
\mu_1&=n\left(1-F_{n-1}\right)=nG_1\label{eqmom1}\\
\begin{split}
\mu_2&=n\left[ n - (2n-1) F_{n-1} + (n-1)F_{n-2}\right]\\
&=n\left[(n-1)G_2 + G_1\right].\label{eqmom2}
\end{split}
\end{align}
The variance evaluates to
\begin{equation}
\mu_2 - \mu_1^2=n\left[ F_{n-1} - n F_{n-1}^2 + (n-1) F_{n-2}\right],
\end{equation}
whence in particular:
\begin{equation}\label{ineq2ndmoment}
F_{n-2}\ge (n-1)^{-1}\left( n F_{n-1} -1 \right) F_{n-1}.
\end{equation}

Two examples are of primary importance in our context:

\textbf{Example 1:} The binomial distribution $p_m=(1-q)^mq^{n-m}$ with
parameter $1-q$. Here we have:
\begin{eqnarray}
f(t)&=&\left(q+(1-q)t\right)^n\\
F(t)&=&(q+t)^n\\
G(t)&=&(1-q+t)^n\\
F_m&=&q^{n-m}\label{eqbinmom}\\
G_m&=&(1-q)^m\label{eqbinmom2}\\
\mu_1&=&n(1-q)\\
\mu_2&=&n(1-q)\left[n(1-q)+q\right]\\
\mu_2-\mu_1^2&=&nq(1-q).\label{bivar}
\end{eqnarray}

\textbf{Example 2:} Fixed weight. Here only bit patterns with a fixed number
$w$ of $1$-bits are permitted. We get:
\begin{eqnarray}
p_m&=&\begin{cases}
{\binom{n}{w}}^{-1}&m=w\\
0&m\neq w
\end{cases}\\
f(t)&=&t^w\label{eq81}\\
F(t)&=&(1+t)^{n-w}t^w\\
G(t)&=&(1+t)^wt^{n-w}\\
F_m&=&\begin{cases}
\frac{\binom{n-w}{n-m}}{\binom{n}{m}}&m\ge w\label{eq83}\\
0&m<w
\end{cases}\\
G_m&=&\begin{cases}
\frac{\binom{w}{m}}{\binom{n}{m}}&m\leq w\\
0&m>w
\end{cases}\\
\mu_m&=&w^m\\
\mu_2-\mu_1^2&=&0.
\end{eqnarray}
Hardly surprising, inequality~(\ref{ineq2ndmoment}) is sharp for this
distribution.

\section{Uniform code word generation}\label{secfwc}
Denoting the target distribution's coefficients by $\check{F}_a$
equation~(\ref{eqtardis}) can be written as:
\begin{equation}\label{eqtardis2}
\check{F}_a=
\sum_{\beta\in\dot{I}^N}P(\beta)
\prod_{j\in\beta^{-1}(1)}F^{(j)}_a,
\end{equation}
where the coefficients $F^{(j)}_a$ describe the random experiment used for
generating the $j$-th code word. If we take these independent of $j$,
(\ref{eqtardis2}) simplifies considerably. Introducing the source bits
probability generating function
\begin{equation}
\Pi(t):=\sum_{\beta\in\dot{I}^N}P(\beta)t^{\#\beta^{-1}(1)},\label{eqdefgnf}
\end{equation}
we can formulate a simple but significant theorem:

\begin{theorem}\label{thmmain}
If the code word generation is performed uniformly with coefficients
$F_a$ and $\Pi$ is the generating function of
the source pattern space, then the target bit distribution
is given by $\check{F}_m=\Pi\left(F_m\right)$.
\end{theorem}

This theorem is the pivot enabling us to compute the target distribution when
source and code distribution are known.
The source bit distribution acts as transformation turning the code bit
distribution into the target bit distribution. The target distribution depends
linearly on the source distribution but in a complex way on the code
distribution.

We can immediately derive the first two moments of the target bit distribution
from equations~(\ref{eqmom1}) and (\ref{eqmom2}):
\begin{align}
\check{\mu}_1&=n\left[1-\Pi\left(F_{n-1}\right)\right]\label{eqcexp}\\
\check{\mu}_2&=n\left[ n - (2n-1) \Pi\left(F_{n-1}\right)
+ (n-1)\Pi\left(F_{n-2}\right)\right]\\
\check{\mu}_2 - \check{\mu}_1^2&=n\left[ \Pi\left(F_{n-1}\right)
- n \Pi\left(F_{n-1}\right)^2 +
(n-1) \Pi\left(F_{n-2}\right)\right]\label{eqcvar}
\end{align}

\textbf{Example 3:} Let's consider fixed weight source words of weight $r$ and
binomially generated code words with parameter $q$. 
From (\ref{eq81}) we can read off the transformation function
$\pi(t)=t^r$, and by (\ref{eqbinmom}) the coefficients are given by
$F_m=q^{n-m}$. Now our theorem implies
$\check{F}_m=q^{r(n-m)}$, i.e., the target bits are binomially distributed
with parameter $q^r$. Then (\ref{eqbinmom2}) immediately implies
$\check{G}_m=\left(1-q^r\right)^m$,
and by (\ref{bivar}) the variance equals
$\check{\mu}_2 - \check{\mu}_1^2=nq^r\left(1-q^r\right)$.
This case is simple because the individual target bits
are independently distributed, which is not true in general.

By definition $\check{G}_m$ equals the expected relative number of candidates
matching a test pattern of length $m$. They are determined by statistics
without reference to the original content, hence they are considered ``false
drops''\footnote{This idiomatic expression is historic and derives from the
application to punched cards.} and their number shall be minimized. Pursuing
our example further, we assume that the test pattern is obtained by submitting
a query source string of weight $s$ to the same superimposed coding
process. Any pattern of weight $m$ in the target query space will occur with
probability
$\tilde{p}_m=\left(1-q^s\right)^mq^{s(n-m)}$ and we can expect a proportion of
$\vartheta=\sum_{m=0}^n\binom{n}{m}\check{G}_m\tilde{p}_m=
\left[\left(1-q^r\right)\left(1-q^s\right)+q^s\right]^n$ random hits.
$\vartheta$ is minimal if we choose $q^s=\frac{r}{r+s}$ with
$\vartheta=\left[1-\left(\frac{r}{r+s}\right)^{\frac{r}{s}}
\frac{s}{r+s}\right]^n$.
For $r\gg s$ we have $q^r=\left(\frac{r}{r+s}\right)^{\frac{r}{s}}=
\left(1-\frac{s}{r+s}\right)^{\frac{r}{s}}
\approx e^{-\frac{r}{r+s}}\approx e^{-1}$ and
$\vartheta=\left[1-\frac{1}{e}
\left(1-q^s\right)\right]^n
=\left[1-\frac{1}{e}\left(1-e^{-\frac{s}{r}}\right)\right]^n$,
hence $\ln\vartheta\approx -\frac{ns}{er}$.
If we do not want the relative number of false drops to exceed a proportion of
$\vartheta_{\text{max}}$ but have to allow for query words of length at least
$s_{\text{min}}$, then we must choose our target bit patterns of length
$n\gtrsim\frac{re}{s_{\text{min}}}\left|\ln\vartheta_{\text{max}}\right|$.
Notice that the original bitstring length $N$ does not enter at all, just the
number of $1$-bits $r$ is relevant.

\textbf{Example 4:}
Our example above was chosen for its simplicity; we don't recommend using
binomially distributed code words in practice. Fixed weight code words 
of weight $w$ can be
expected to exhibit a much more robust behavior. Because of
equations~(\ref{eqcexp}) and (\ref{eqcvar}) and the monotonicity of probability
generating functions fixed weight code words will produce minimal variance in
the target distribution if the expectation is prescribed.
Both cases may be directly compared if the
parameter of the binomial distribution is chosen as
$q=1-\frac{w}{n}$, because by (\ref{eq83})
$F_{n-1}=\binom{n-w}{1}/\binom{n}{n-1}=\frac{n-w}{n}=q$.
Furthermore $F_{n-2}=\binom{n-w}{2}/\binom{n}{n-2}
=\frac{(n-w)(n-w-1)}{n(n-1)}=q\frac{nq-1}{n-1}$ and
$\check{F}_{n-2}=q^r\left(\frac{nq-1}{n-1}\right)^r$.
The variance
\begin{align}
\text{var}&=\check{\mu}_2-\check{\mu}_1^2=n\left\{ q^r -nq^{2r}
+(n-1)q^r\left(\frac{nq-1}{n-1}\right)^r\right\}\\
&=n\left\{q^r-q^{2r}-(n-1)\left[q^{2r}-q^r\left(q-\frac{1-q}{n-1}
\right)^r\right]\right\}
\end{align}
can be computed asymptotically
for large $n$ by developing the rightmost bracket in a power series
$\left(q-\frac{1-q}{n-1}\right)^r=q^r-rq^{r-1}\frac{1-q}{n-1}\pm\cdots$, where
all higher powers may be safely discarded:
\begin{equation}
\text{var}\approx
nq^r\left(1-q^r\right)\left[1- r(1-q)
\frac{q^{r-1}}{1-q^r}\right],
\end{equation}
the first factor being equal to the value in the binomial case. With suitable
parameters the factor in square brackets may be quite small, i.e., the fixed
weight variance may be negligible while the binomial variance is not.

We reconsider example~3 with fixed weight code words. In \cite{Rob79} it is
shown\footnote{As a matter of fact the equation is derived by using the
binomial case as approximation and observing that target query weights have
small variance.}
$\vartheta\approx\left(1-q^r\right)^{n\left(1-q^s\right)}$. The minimum value
is attained for $q^r\approx\frac{1}{2}$ with
$\ln\vartheta\approx-\frac{ns}{r}\left(\ln2\right)^2$.
This value is slightly
better than in example~3. We have
$\text{var}\approx\frac{ns^2}{r^2}\left(\ln2\right)^2$, which is by a factor
$\frac{s}{r}\ln2$ smaller than the binomial case.

\section{Adapting to non uniform but independent
source bit distributions}\label{secNuInd}
If the individual source bits are independent and the $i$-th bit is turned on
with probability $p_i$ the probability of a given source pattern
$\beta\in\dot{I}^N$ is
\begin{eqnarray}
P(\beta)&=&
\prod_{i\in\beta^{-1}(1)}p_i
\prod_{i\not\in\beta^{-1}(1)}\left(1-p_i\right)\label{eqindprob}\\
\check{F}_a&=&\prod_{j=1}^N\left(p_jF^{(j)}_a+1-p_j\right)\label{inhomind}
\end{eqnarray}
(\ref{inhomind}) is obtained by inserting (\ref{eqindprob}) into
(\ref{eqtardis2}) and distributively collecting terms. In our selection of
optimal code word distributions we let ourselves be guided by what we learned
in section~\ref{secfwc}: We set $\check{F}_{n-1}=\frac{1}{2}$ and minimize
$\check{F}_{n-2}$, observing that by equation~(\ref{eqmom2}) this is
equivalent to minimization of the second moment and hence of the
variance. By inequality~(\ref{ineq2ndmoment})
$F^{(j)}_{n-2}\ge(n-1)^{-1}\left(nF^{(j)}_{n-1}-1\right)F^{(j)}_{n-1}$,
the lower bound being attained for fixed weight code words. Substituting
\begin{equation}
u_j:=p_jF^{(j)}_{n-1}+1-p_j
\end{equation}
we have to solve the minimization problem
\begin{align}
\begin{split}
\check{F}_{n-2}&=\frac{n^N}{(n-1)^N\prod_jp_j}\cdot\\
&\cdot\prod_{j=1}^N\left\{\left(u_j-\nu_j\left(1-p_j\right)\right)^2
+\omega_j p_j\left(1-p_j\right)\right\}
=\text{min}\label{eqmin}
\end{split}\\
\intertext{under the constraint}
\frac{1}{2}&=\check{F}_{n-1}=\prod_{j=1}^Nu_j\\
\intertext{in the domain}
&1-\frac{n-1}{n}p_j\le u_j\le 1\label{eqdombnd}\\
\intertext{with}
\nu_j&:= 
\left(1-p_j\right)^{-1}\left[1-\left(1-\frac{1}{2n}\right)p_j\right]\\
\begin{split}
\omega_j&:=\left(1-p_j\right)^{-1}
\left[1-\frac{1}{n}-\left(1-\frac{1}{2n}\right)^2p_j\right]\\
&=p_j^{-1}\left[1-\nu_j^2\left(1-p_j\right)\right].
\end{split}
\end{align}
Observe that the terms $\nu_j$ and $\omega_j$
introduced for abbreviation are very close to
$1$, in particular $\omega_j>0$. Ignoring the restrictions (\ref{eqdombnd})
temporarily, we see that (\ref{eqmin}) tends to $+\infty$ if at least one of
the coordinates $u_j$ does, therefore (\ref{eqmin}) must have an absolute
minimum. 
We are going to locate it using a Lagrange multiplier
and will then have to check conditions (\ref{eqdombnd}). Thus
\begin{align}
0&=\frac{\partial}{\partial u_j}\left(
\ln\check{F}_{n-2}-\lambda\sum_{j=1}^N\ln u_j\right)\\
2\left[u_j-\nu_j\left(1-p_j\right)\right]u_j&=
\lambda\left[u_j^2-2\nu_j\left(1-p_j\right)u_j
+1-p_j\right]\label{eqquadr}\\
\prod_{j=1}^N\frac{u_j-\nu_j\left(1-p_j\right)}{p_j}&=
2\left[\frac{\lambda}{2}\left(1-\frac{1}{n}\right)\right]^N\check{F}_{n-2}.
\end{align}
We see that $\frac{\lambda}{2}$ will be approximately the geometric mean of
the quantities $F^{(j)}_{n-1}$ and hence close to $1$.
Expanding
$u_j=\alpha+\beta(\lambda-2)+\gamma\left(\nu_j-1\right)\pm\cdots$
up to linear order in the small quantities $\lambda-2$ and
$\nu_j-1$ and substituting into (\ref{eqquadr}) leads to
\begin{eqnarray}
\lambda-2&\approx& \frac{1}{n}
-\frac{2\ln2}{\sum_{k=1}^N\frac{p_k}{1-p_k}}\\
u_j&\approx&1-\frac{p_j}{1-p_j}
\frac{\ln2}{\sum_{k=1}^N\frac{p_k}{1-p_k}}\\
F^{(j)}_{n-1}&\approx&1-\frac{1}{1-p_j}
\frac{\ln2}{\sum_{k=1}^N\frac{p_k}{1-p_k}}.
\end{eqnarray}
Summarizing:

\begin{theorem}\label{thmindnonuni}
An optimal target distribution is obtained by choosing fixed weight code words
of weight
$\frac{n}{1-p_j}
\frac{\ln2}{\sum_{k=1}^N\frac{p_k}{1-p_k}}$
for encoding of bit $j$.
\end{theorem}

If the probabilities $p_j$ are actually independent of $j$ then this coincides
with section~\ref{secfwc}.

\section{The general case}\label{secgencas}
There remains the question what to do if the source bit distributions are
neither uniform nor independent. We may take a clue from
theorem~\ref{thmindnonuni}: as long as
the individual bit probabilities $p_j$ are not too large, say $<1/2$, then the
code word lengths recommended there do not vary significantly. We may try one
and the same code distribution for all source bits and thus place ourselves in
the situation of theorem~\ref{thmmain}. The generating function defined in
equation~(\ref{eqdefgnf}) is the same that would be obtained from an isotropic
bit distribution with
\begin{equation}
p_m=\binom{N}{m}^{-1}\sum_{\beta\in\dot{I}^N,\#\beta^{-1}(1)=m}P(\beta).
\end{equation}
Choosing fixed length code words of weight $n(1-q)$
for a suitable parameter $q$
theorem~\ref{thmmain}
tells us that the target bits will have an expected weight of $\Pi(q)$ and we
want to arrange for $\Pi(q)=1-\check{G}_1=\frac{1}{2}$.
Substituting $q=e^{-\varepsilon}$ with $0<\varepsilon\ll1$ we derive from
(\ref{eqmomgen}):
\begin{equation}
\check{G}_1=1-\Pi\left(e^{-\varepsilon}\right)=
\sum_{m=1}^\infty (-1)^{m+1}\frac{\mu_m}{m!}\varepsilon^m.
\end{equation}
This equation is quite suitable for practical application, because the power
series is fast converging and the lower moments of the source bit distribution
are easily evaluated.
We can solve for $\varepsilon$:
\begin{equation}
\varepsilon=
\frac{1}{\mu_1}\check{G}_1+
\frac{\mu_2}{2\mu_1^3}\check{G}_1^2+
\frac{3\mu_2^2-\mu_1\mu_3}{6\mu_1^5}\check{G}_1^3\pm\cdots
\end{equation}
Convergence of this power series is again fast enough to use the partial sum
above as practical estimate.

Notice that in case of source words of fixed weight $r$
we have $\mu_m=r^m$ and we recover section~\ref{secfwc} exactly.

\bibliography{distsupcodref}
\bibliographystyle{IEEEtran}

\end{document}